\documentclass[12pt]{article}
\usepackage[english]{babel}
\usepackage{amsfonts}
\begin{document}

\begin{center}
\rightline{NPI MSU 2003-6/719} \rightline{February 2003}
\vspace{1cm}

{\LARGE\bf Is there the radion in the RS2 model ?} \vspace{4mm}

{Mikhail N. Smolyakov$^{a}$, Igor P. Volobuev$^{b}$\\ \vspace{2mm}
$^a$Physics Department, Moscow State University
\\ 119992 Moscow, Russia \\ $^{b}$Institute of Nuclear Physics,
Moscow State University \\ 119992  Moscow, Russia }

\end{center}

\begin{abstract}{We analyse the physical boundary conditions at infinity
for metric fluctuations and  gauge functions in the RS2 model with matter
on the brane. We argue that due to these boundary conditions the radion
field cannot be gauged out in this case. Thus, it represents a physical
degree of freedom of the model.}
\end{abstract}

\section{Introduction}
The RS2 model \cite{RS2}  is based on the solution for the background
metric, which was obtained from the solution for the background metric of
the Randall-Sundrum model with two branes \cite{RS1} by pushing the
negative tension brane to infinity. The model  describes  gravity in a
five-dimensional space-time $E$ with one brane embedded into it. We denote
the coordinates  in $E$ by $\{ x^M\} \equiv \{x^{\mu},x^4\}$, $M=
0,1,2,3,4, \, \mu=0,1,2,3$, the coordinate $x^4 \equiv y$ parameterizing
the fifth dimension, which is infinite. The brane is located at $y=0$, and
all the fields possess a symmetry  under the reflection $y\leftrightarrow-y$,
which is inherited from the RS1 model. Explicitly,  it reads
\begin{eqnarray}
\label{orbifoldsym}
 g_{\mu \nu}(x,- y)=  g_{\mu \nu}(x,  y), \\
 \nonumber
  g_{\mu 4}(x,- y)= - g_{\mu 4}(x,  y), \\ \nonumber
   g_{44}(x,- y)=  g_{44}(x,  y).
\end{eqnarray}
The meaning  of this symmetry can  be  easily understood even without
referring to the RS1 model: if matter is localized on the brane, all the
physical fields should possess a symmetry  under the reflection in the
brane.

It is a common knowledge that among the degrees of freedom of the
RS1 model there is a massless scalar field, called the radion
and describing the oscillations of the branes with respect to each other.
At the first glance, it seems to be very likely that this degree of freedom
should drop from the model,  if one brane is pushed to infinity. In fact, this
assumption was made in papers \cite{RS2,GarTan,RKatz}, where it was noted that
the 44-component of the metric fluctuations, which corresponds to the scalar
mode, could be gauged out. However, in this gauge the brane is located not at $y=0$,
but at $y=\xi(x)$. Obviously, this "bent-brane" formulation destroys the reflection
symmetry (\ref{orbifoldsym}), which makes the approach based on this gauge inconsistent;
this fact was noted in \cite{AIMVV}. In paper \cite{Kakushadze} it was observed that
gauging out the radion field in the straight brane formulation with matter on the brane
leads to unphysical solutions, which diverge at infinity. Thus,  gauging
out the radion field resulted in some discrepancies, and it looks as if this field
were of particular importance in the RS2 model. In the present paper we are going to
study the role of the radion in the RS2 model more thoroughly. We begin with briefly
discussing the main features of the RS2 model.

The action of the model is
\begin{equation}\label{actionRS}
 S = S_g + S_{brane},
\end{equation}
where $S_g$ and $S_{brane}$ are given by
\begin{eqnarray}\label{actionsRS}
S_g&=& \frac{1}{16 \pi \hat G} \int_E
\left(R-\Lambda\right)\sqrt{-g}\, d^{4}x dy,\\ \nonumber
 S_{brane}&=& V \int_E \sqrt{-\tilde g} \delta(y) d^{4}x dy.\\ \nonumber
\end{eqnarray}
{Here $\tilde g_{\mu\nu}$ is the induced metric on the brane and $V$ is the
brane tension.} We also note that the signature of the metric $g_{MN}$ is
chosen to be $(-,+,+,+,+)$.

The Randall-Sundrum solution for the  metric is {given by}
\begin{equation}\label{metricrs}
ds^2=  g_{MN} d{x}^M d{x}^N = e^{2\sigma(y)} \eta_{\mu\nu}
{dx^\mu dx^\nu} +
  dy^2,
\end{equation}
where $\eta_{\mu\nu}$ is the Minkowski metric and {the function}
$\sigma(y) = -k|y|$. The parameter  $k$ is positive and has the
dimension of mass; the parameters $\Lambda$ and $ V$ are {related
to it as follows:} $$ \Lambda = -12 k^2, \quad V = -\frac{3k}{4\pi
\hat G}. $$ We see that the brane has a positive energy density.
The function $\sigma$ has the properties
\begin{equation}\label{sigma}
  \partial_4 \sigma = -k\, sign(y), \quad \frac{\partial^2 \sigma}{\partial
  {y}^2} =-2k\delta(y).
\end{equation}
Here and in the sequel $\partial_4 \equiv \frac{\partial}{\partial y}$.

We denote $\hat \kappa = \sqrt{16 \pi \hat G}$, where $\hat G$ is the
five-dimensional gravitational constant, and parameterize the metric
$g_{MN}$ as
\begin{equation}\label{metricpar}
  g_{MN} = \gamma_{MN} + \hat \kappa h_{MN},
\end{equation}
$h_{MN}$ being the metric fluctuations. Substituting this
parameterization into (\ref{actionRS}) and retaining the terms of
the zeroth order in $\hat \kappa$, we can get the second variation
action of this model. In \cite{BKSV} the second
variation action for the  RS1 model was obtained, and we can apply this
result to the RS2 model just  by changing the definition of $\sigma(y)$ and
of its derivatives.

The action is invariant under the gauge transformations
\begin{eqnarray}\label{gaugetrRS}
h'_{MN}(x,y) = h_{MN}(x,y) -(\nabla_M\xi_N(x,y) + \nabla_N\xi_M(x,y) ),
\end{eqnarray}
where $\nabla_M$ is the covariant derivative with respect to the
background metric $\gamma_{MN}$, and the functions $\xi^M(x,y)$
satisfy the symmetry conditions
\begin{eqnarray}\label{orbifoldsym1}
\xi^{\mu}\left(x,-y\right)&=&\xi^{\mu}\left(x,y\right)\\
\nonumber
\xi^{4}\left(x,-y\right)&=&-\xi^{4}\left(x,y\right).
\nonumber
\end{eqnarray}
Equations (\ref{gaugetrRS}) can be rewritten in a more useful
component form as follows:
\begin{eqnarray}\label{gaugetrRSuseful}
h'_{\mu\nu}\left(x\right)&=&h_{\mu\nu}\left(x\right)-\left(
\partial_{\mu}\xi_{\nu} +\partial_{\nu}\xi_{\mu}+2\gamma_{\mu\nu}\partial_{4}\sigma\xi_{4}
\right),\\ \label{gaugetrRSuseful2}
h'_{\mu4}\left(x\right)&=&h_{\mu4}\left(x\right)-\left(
\partial_{\mu}\xi_{4} +\partial_{4}\xi_{\mu}-2\partial_{4}\sigma\xi_{\mu}
\right),\\ \label{gaugetrRSuseful3}
h'_{44}\left(x\right)&=&h_{44}\left(x\right)-2\partial_{4}\xi_{4}.
\end{eqnarray}

\section{Gauge conditions and equations of motion for \\ linearized gravity}
Now let us discuss the gauge conditions  and equations of motion
for linearized gravity in the presence of matter on the brane.
The interaction with matter on the brane has the standard form
\begin{equation}\label{interaction}
 \frac{\hat \kappa}{2} \int_{brane} h^{\mu\nu}(x,0) t_{\mu\nu}(x) dx,
\end{equation}
$t_{\mu\nu}(x)$ denoting the energy-momentum tensor of the matter.

First, we would like to emphasize that in general all fluctuations of
metric must satisfy the physical boundary conditions at  $y\to\pm\infty$,
$ x^i\to\pm\infty \quad (i=1,2,3)$, i.e. vanish at spatial infinity. This
is a reasonable assumption - for example, the $h_{00}$-component is
associated with  Newton's potential, which must vanish at infinity (for the
matter, which is localized in some finite domain). Below we will show that
the fields of certain exact solutions (for example, with point-like matter
sources) do satisfy these boundary conditions.

Obviously, the gauge functions  $\xi^M(x,y)$ must be finite everywhere in
$E$. This means that $$
\xi_{\mu}=e^{2\sigma}\eta_{\mu\nu}\xi^{\nu}|_{y\to\pm\infty} \to 0. $$ The
situation with $\xi_4$ is more complicated, since
$\xi_4=g_{44}\xi^4=\xi^4$. Let us consider equation
(\ref{gaugetrRSuseful2}). It follows from this equation  that if $\xi_4$
does not depend on four-dimensional coordinates $x$, we can satisfy the
physical boundary condition for the field $h_{\mu 4}$ without requiring
$\xi_4$ to vanish at $y\to\pm\infty$. But in this case the last term in the
r.h.s. of (\ref{gaugetrRSuseful}) does not satisfy the physical boundary
condition for the field $h_{\mu\nu}$ at $ x^i\to\pm\infty$. This means that
$\xi_4$ must be $x$-dependent and vanish at $ x^i\to\pm\infty$. At the same
time the form of the second term in the r.h.s. of (\ref{gaugetrRSuseful2})
shows that $\xi_4$ must vanish at $y\to\pm\infty$ to satisfy the physical
boundary conditions for the field $h_{\mu 4}$. Thus, we have the following
boundary conditions for $\xi_M$:
\begin{equation}
\xi_M|_{ x^i,y\to\pm\infty}\to 0
\end{equation}

Now let us examine the equations of motion. They look as follows:

1) $\mu\nu$-component
\begin{eqnarray}\label{mu-nu}
 & &\frac{1}{2}\left(\partial_\rho \partial^\rho h_{\mu\nu}-
\partial_\mu \partial^\rho
h_{\rho\nu}-\partial_\nu \partial^\rho h_{\rho\mu} +
\partial_4\partial_4 h_{\mu\nu}\right)- \\
\nonumber &-& 2k^2h_{\mu\nu}+\frac{1}{2}\partial_\mu
\partial_\nu\tilde h+ \frac{1}{2}\partial_\mu \partial_\nu
h_{44}-\partial_4\sigma(\partial_{\mu}h_{\nu
4}+\partial_{\nu}h_{\mu 4})
-\frac{1}{2}\,\partial_4(\partial_{\mu}h_{\nu
4}+\partial_{\nu}h_{\mu 4})+
\\ \nonumber &+& \frac{1}{2} \gamma_{\mu\nu}\Biggl(\partial^\rho
\partial^\sigma h_{\rho\sigma}-\partial_\rho \partial^\rho \tilde
h - \partial_4\partial_4 \tilde h -4\partial_4\sigma
\partial_4 \tilde h - \partial_\rho \partial^\rho h_{44} + 12 k^2 h_{44}
+ 3\partial_4\sigma\partial_4 h_{44}+ \\ \nonumber &+&
2\partial^{\rho}\partial_4 h_{\rho
4}+4\,\partial_4\sigma\partial^{\rho}h_{\rho 4} \Biggr)
 +\left(2k  h_{\mu\nu} - 3k\gamma_{\mu\nu}h_{44}
\right)\delta(y) = -\frac{\hat \kappa}{2} t_{\mu\nu}(x)\delta(y),
\end{eqnarray}

 2) $\mu 4$-component
\begin{equation}\label{mu-4}
\partial_4 ( \partial_\mu \tilde h - \partial^\nu  h_{\mu\nu})-
3\partial_4 \sigma \partial_\mu h_{44} -
\partial^{\nu}\left(\partial_{\mu}h_{\nu 4}-\partial_{\nu}h_{\mu 4}\right)= 0,
\end{equation}

 3) $4 4$-component
\begin{equation}\label{4-4}
\frac{1}{2}(\partial^\mu \partial^\nu  h_{\mu\nu} - \partial_\mu
\partial^\mu \tilde h ) - \frac{3}{2}\partial_4 \sigma \partial_4 \tilde h
+ 6 k^2 h_{44} +3\partial_4\sigma\partial^{\mu}h_{\mu 4}=0,
\end{equation}
where $\tilde h=\gamma^{\mu\nu}h_{\mu\nu}$.

In what
follows, we will also use an auxiliary equation, which is obtained
by multiplying the equation for $44$-component by 2 and
subtracting it from the contracted equation for
$\mu\nu$-component. This equation contains $\tilde h$, $h_{\mu 4}$
and $h_{44}$ only and has the form:
\begin{equation}\label{contracted-44}
\partial_4\left(e^{2\sigma}\partial_4 \tilde h \right)
-4\partial_4\left(e^{2\sigma}\partial_4 \sigma
h_{44}\right)-2\partial_4(e^{2\sigma}\partial^{\mu}h_{\mu 4})+
\Box h_{44} = \frac{\hat \kappa}{3}\, t_{\mu}^{\mu}(x)\delta(y),
\end{equation}
where $\Box=\eta^{\mu\nu}\partial_{\mu}\partial_{\nu}$. By
integrating this equation in the limits $(-\infty,\infty)$ and
using the physical boundary conditions for the fields $h_{\mu\nu}$,
$h_{\mu 4}$ and $h_{44}$, we  find that the function $\varphi(x)$, defined by
\begin{equation}
\int_{-\infty}^\infty h_{44}(y)\, dy = \varphi(x),
\end{equation}
is not equal to zero and satisfies the equation
\begin{equation}\label{h44t}
\Box\varphi(x) = \frac{\hat
\kappa}{3}\, \eta^{\mu\nu}t_{\mu\nu}(x)\equiv\frac{\hat \kappa}{3}\,
t(x).
\end{equation}
Obviously, the admissible gauge transformations do not alter the physical boundary
conditions for the metric fluctuations, and therefore equation (\ref{h44t}) holds in any
gauge. This is an equation for a four-dimensional scalar field, which coincides with
the equation for the radion field in the RS1 model with matter on the positive tension
brane \cite{SV}.

Thus, we arrive at the corollary that the radion field  cannot be  gauged
out in the RS2 model, because otherwise the equations  of motion for
linearized gravity become inconsistent. In other words,  this means that
the gauge functions $\xi_4$, corresponding to the gauge $h_{44}= 0$, do not
satisfy the boundary conditions at $y\to\pm\infty$, which is a good check
of the  consistency of our approach. In fact, this was noted in
\cite{Kakushadze}. It was shown there that the solutions for the linearized
gravity in the absence of the radion are unphysical, i.e. they diverge at
$y=\pm\infty$.

We will use the following form of $\xi_4$ to impose an appropriate gauge on the
field $h_{44}$:
\begin{equation}\label{xi4}
\xi_4(x,y)=\frac{1}{4}\int_{-y}^y h_{44} (x,y') dy' -\frac{1}{4C}
\int_{-y}^y F(y') dy' \int_{-\infty}^{\infty} h_{44} (x,y') dy',
\end{equation}
where $F|_{y\to \pm\infty}=0$ and
\begin{equation}
C=\int_{-\infty}^{\infty} F(y) dy.
\end{equation}
Note that $\xi_4$ satisfies the symmetry and the boundary conditions.
With the help of (\ref{xi4}) we can pass to the gauge, in which
\begin{equation}\label{gaugephi}
h_{44}(x,y)=F(y)\phi(x),
\end{equation}
where
\begin{equation}
\phi(x)=\frac{1}{C}\int_{-\infty}^{\infty} h_{44} (x,y) dy
\end{equation}
and depends on $x$ only. It turns out to be convenient to choose
$F(y)=e^{2\sigma}=e^{-2k|y|}$. Obviously, the field $h_{44}$ satisfies the
symmetry and the physical boundary conditions in this gauge. Moreover, we
have no residual gauge transformations with $\xi_4$. We also note that
since $\xi_4(x,0)=0$, the brane remains straight in this gauge, i.e. we
{\it do not} use the bent-brane formulation \cite{GarTan,RKatz}, which
allegedly destroys the structure of the model (this problem was discussed
in \cite{AIMVV}).

We would like to note that the gauge choice of the type (\ref{gaugephi})
with an arbitrary finite even function $F(y)$ can be used it the RS1 model
as well. For example, a gauge with $h_{44}(x,y)\sim e^{2k|y|}\,\phi(x)$ was
used in \cite{Toharia}.

Now let us discuss the  gauge condition for the field $A_{\mu}=h_{\mu4}$.
Let us take the gauge function $\xi_{\mu}(x,y)$ in the following form:
\begin{equation}\label{gaugetrMuNu}
\xi_{\mu}(x,y)=e^{2\sigma}\int_{-\infty}^{y}e^{-2\sigma}
A_{\mu}\left(x^{\nu},y\,'\right)\,dy'.
\end{equation}
Of course, this  definition makes sense, if the field $A_{\mu}$ is such
that the integral in (\ref{gaugetrMuNu}) is well convergent to provide  an
acceptable ($\sim e^{2\sigma}$) decrease of $\xi_{\mu}$. One can easily see
that due to the symmetry $A_{\mu}(x,-y)=-A_{\mu}(x,y)$ (see
(\ref{orbifoldsym})), $\xi_{\mu}(x,y)$ satisfies the symmetry condition
$\xi_{\mu}(x,-y)=\xi_{\mu}(x,y)$. Moreover, it is easy to see that
$\xi_{\mu}(x,y)|_{y\to\pm\infty}\to 0$, at least in the sense of the
principal value of the integral in eq. (\ref{gaugetrMuNu}) (again due to
the symmetry of $A_{\mu}$). Finally, it is not difficult
to check that the gauge transformation with $\xi_{\mu}$ given by
(\ref{gaugetrMuNu}) gauges the field $A_{\mu}$ out.

We think that this formal argumentation can be used in favor of the
possibility to make the $A_{\mu}$-field vanish  everywhere. Moreover, with
a different motivation, an expression similar to  (\ref{gaugetrMuNu}) was
assumed to be well defined in \cite{AIMVV}. Anyway, in all the papers
concerning the RS2 model it is universally recognized that the field
$A_{\mu}$ can be gauged away (see, for example, \cite{Kakushadze}). Thus,
we also adhere to this opinion. As we will see later, equations of motion
can be solved exactly in the gauge $A_{\mu}= 0$ (see also
\cite{Kakushadze}).

After this gauge fixing we are still left with residual gauge
transformations of the form
\begin{equation}\label{remgaugetr}
\partial_{4}\left(e^{-2\sigma}\xi_{\mu}\right)=0.
\end{equation}

Now we are ready to solve equations of motion in the gauge
\begin{eqnarray}\label{unitgauge}
h_{\mu4}(x,y)= 0, \\ \nonumber
h_{44}(x,y)=e^{2\sigma}\phi(x).
\end{eqnarray}

\section{Solution of the equations of motion}
The substitution, which allows us to solve equations of motion in the gauge
(\ref{unitgauge}), has the form
\begin{equation}\label{substitution}
h_{\mu\nu}=b_{\mu\nu}+\frac{1}{2}\,e^{2\sigma}\gamma_{\mu\nu}\phi-
\frac{1}{2k^2}\,\sigma \,
e^{2\sigma}\partial_{\mu}\partial_{\nu}\phi.
\end{equation}
Note that if $b_{\mu\nu}|_{y\pm\to\infty}\to 0$, then
$h_{\mu\nu}|_{y\pm\to\infty}\to 0$. Substituting (\ref{substitution}) into
(\ref{mu-4}), (\ref{4-4}), (\ref{contracted-44})  and using the notation
$\tilde b = \gamma^{\mu\nu}b_{\mu\nu}$, similar to the one utilized in
(\ref{mu-4}), (\ref{4-4}), we get
\begin{equation}\label{mu-4-1}
\partial_4 ( \partial_\mu \tilde b - \partial^\nu  b_{\mu\nu})
= 0,
\end{equation}
\begin{equation}\label{4-4-1}
(\partial^\mu \partial^\nu  b_{\mu\nu} - \partial_\mu
\partial^\mu \tilde b) - 3\partial_4 \sigma \partial_4 \tilde b =0,
\end{equation}
\begin{equation}\label{contracted-44-1}
\partial_4\left(e^{2\sigma}\partial_4 \tilde b \right)+\frac{1}{k}
\,\Box\,\phi \delta(y)  = \frac{\hat \kappa}{3}\,
t_{\mu}^{\mu}(x)\delta(y).
\end{equation}
Integrating (\ref{contracted-44-1}) in the limits $(-\infty,\infty)$ and
using the physical boundary conditions for the field $b_{\mu\nu}$, we get
\begin{equation}\label{phi}
\Box\,\phi = \frac{\hat \kappa k}{3} t.
\end{equation}
This mean that
\begin{equation}
\partial_4 \tilde b =B(x) \, e^{-2\sigma},
\end{equation}
where $B(x)$ is some function of $x$ only. Using the symmetry conditions
(\ref{orbifoldsym}), we obtain $B(x)\equiv 0$.

Recall that we have at our disposal the gauge transformations satisfying
(\ref{remgaugetr}). With the help of these transformations, we can impose
the gauge
\begin{equation}\label{T}
\tilde b = b =0,
\end{equation}
where $b=\eta^{\mu\nu}b_{\mu\nu}$. It is easy to see that there
remain gauge transformations parameterized by $\xi_\mu =
e^{2\sigma}\epsilon_\mu(x)$ with $\epsilon_\mu(x)$ satisfying
$\partial^\mu \epsilon_\mu = 0$. Substituting expression (\ref{T})
into (\ref{mu-4-1}) and (\ref{4-4-1}) we arrive at the following
system of relations:
\begin{eqnarray}
\label{system3} \label{equ1}
\partial^\mu \partial^\nu  b_{\mu\nu}& =&0, \\
\label{equ2}
\partial_4 (e^{-2\sigma} \partial^\mu  b_{\mu\nu})&=&0,
\end{eqnarray}
where indices are raised with flat Minkowski metric $\eta^{\mu\nu}$. The
remaining gauge transformations are sufficient to impose the condition
\begin{equation}\label{gaugecond}
 \partial^\mu  b_{\mu\nu} = 0.
\end{equation}
The conditions (\ref{T}) and (\ref{gaugecond}) define the gauge, which is
usually called the transverse-traceless (TT) gauge. Having imposed this
gauge, we are still left with residual gauge transformations
\begin{equation}
\label{gaugetr} \xi_\mu = e^{2\sigma}\epsilon_\mu(x), \quad
\Box\epsilon_\mu = 0, \quad \partial^\mu \epsilon_\mu = 0,
\end{equation}
which are important for determining the number of degrees of
freedom of the massless mode of $b_{\mu\nu}$.

Substituting (\ref{substitution}) into (\ref{mu-nu}) and using (\ref{T}),
(\ref{gaugecond}) and (\ref{phi}), we  get the well-known equation
\begin{eqnarray}\label{b-mu-mu}
 \frac{1}{2}\left(e^{-2\sigma}\Box\, b_{\mu\nu}
+ \partial_4\partial_4 b_{\mu\nu}\right)- 2k^2b_{\mu\nu}+2k
b_{\mu\nu}\delta(y) = \\ \nonumber  = -\frac{\hat
\kappa}{2}\,\delta(y)
\left[t_{\mu\nu}-\frac{1}{3}\left(\eta_{\mu\nu}-
\frac{\partial_{\mu}\partial_{\nu}}{\Box}\right)t\right].
\end{eqnarray}
This equation is identical to the one obtained by Garriga and
Tanaka \cite{GarTan}. It was solved exactly, for example, in
\cite{AIMVV}, and the solution for ordinary (not tachyonic) matter
on the brane looks like
\begin{equation}\label{fourier}
b_{\mu\nu}(x,y)=\frac{1}{(2\pi)^4}\int_{p^2>0} e^{-i\eta_{\mu\nu}p^\mu
x^\nu}\tilde b_{\mu\nu}(p,y)d^4p,
\end{equation}
where for $p^2 = -p_0^2 + \vec p^2 >0$ (which includes the  static case
$p_0=0$)
\begin{eqnarray}\label{pbigger0}
\tilde b_{\mu\nu}(p,y)=\left[\tilde
t_{\mu\nu}-\frac{1}{3}\left(\eta_{\mu\nu}-\frac{p_{\mu}p_{\nu}}{p^2}\right)\tilde
t\right]\frac{\hat\kappa}{2\sqrt{p^2}}
\frac{K_2\left(\frac{\sqrt{p^2}}{k}e^{k|y|}\right)}
{K_1\left(\frac{\sqrt{p^2}}{k}\right)}.
\end{eqnarray}
 We note, that (\ref{pbigger0})
coincides with the corresponding formula for the RS1 model,
obtained in \cite{SV}, in the limit  $R\to\infty$.

Thus, the exact solution for linearized gravity in our gauge is given by
(\ref{substitution}), (\ref{T}), (\ref{gaugecond}), (\ref{phi}),
(\ref{fourier}) and (\ref{pbigger0}). Taking into account, that $\phi$ does
not depend on the extra coordinate $y$ and using (\ref{substitution}),
(\ref{fourier}), (\ref{pbigger0}) one can easy see that with a "good"
energy-momentum tensor $t_{\mu\nu}(x)$ (for example, that of a static
point-like source) fields $h_{\mu\nu}(x,y)$ and $h_{44}(x,y)$ decay to zero
at the spatial infinity.

Now let us examine gravity on  the brane. The fluctuations of the
metric on the brane have the following form
\begin{equation}\label{onbrane}
h_{\mu\nu}(x,0)=b_{\mu\nu}(x,0)+\frac{\hat \kappa
k}{6}\,\eta_{\mu\nu}\Box^{-1}t.
\end{equation}
Using (\ref{b-mu-mu}), one can easy see that equation (\ref{onbrane})
coincides with the solutions for gravity on the brane, obtained in
\cite{GarTan,RKatz,AIMVV,Kakushadze}, although we did not use the
"bent-brane" formulation, which was used in \cite{GarTan,RKatz}.

Finally, we have to answer the question, which was posed in the
title of the paper. Obviously, it amounts to finding the number of
the independent degrees of freedom in the RS2 model. As we have
shown, we cannot completely gauge away the radion field in the
presence of  matter on the brane (see (\ref{h44t})). The situation
is rather different, if there is no matter on the brane. In this
case we deal with  equations for the free fields, possessing
solutions  of the plane wave type, which do not vanish at
infinity. Therefore, we have to drop the physical boundary
conditions for all the components of the metric fluctuations.
Thus, there is  no need for the gauge function $\xi_4$ to decay to
zero at $y\to\pm\infty$ (though the functions $\xi_{\mu}$ must
still decay to zero at $y\to\pm\infty$, because  $\xi^M$ must be
finite everywhere in $E$ and
$\xi_{\mu}=e^{2\sigma}\eta_{\mu\nu}\xi^{\nu}$). It means that the
radion field can be gauged out  and is no more  an independent
degree of freedom of the RS2 model in this case. Nevertheless, the
radion field appears, if we place matter  sources on the brane,
and it allows us to solve consistently the equations of motion.

We can find an analogy to this situation in  electrodynamics. It is a
common knowledge that  longitudinal photons do not appear in  the
asymptotic states (on the mass shell), whereas their contribution is
important in the radiative corrections (off the mass shell). The radion
field is very similar to longitudinal photons: it is absent in the
asymptotic states, but it is absolutely necessary for consistently
describing the interaction off the mass shell.

There is another problem, which  may arise in the case of the absence of
matter on the brane. Since we drop the physical boundary conditions for the
field $A_{\mu}$ ($A_{\mu}|_{y\to\pm\infty}\to 0$), the gauge parameter
$\xi_{\mu}$, defined by formula (\ref{gaugetrMuNu}), may not decay to zero
$\sim e^{2\sigma}$ at infinity. This means that there may be additional
degrees of freedom in the RS2 model. This problem deserves  a more detailed
investigation.

\section{Conclusion}
In the present paper we have studied the boundary conditions for
the metric fluctuations and the gauge functions  in the RS2 model
with and without matter on the branes and  solved  exactly the
equations of motion in the presence of matter on the brane  in a
convenient gauge. The validity of the imposed gauge conditions
was carefully checked. The gauge is very simple and is more
transparent from the physical point of view,  than the gauge used
in \cite{AIMVV}, where the equations for linearized gravity were
solved exactly as well. Another advantage of this gauge choice is
that the  brane remains straight in this case. We have shown
that although the radion is not an independent degree of freedom
of the model, it is indispensable in the case of the presence of
matter on the brane (in this case the radion field cannot be
completely gauged away). The analysis made above is completely
equivalent to the one made in \cite{BKSV,SV}, where linearized
gravity in the RS1 model was treated. We believe that the physically
transparent method, which was used in this paper, is useful for
understanding the general structure of both  Randall-Sundrum models.

\bigskip
{ \large \bf Acknowledgments}
\medskip

The authors are grateful to B.A. Arbuzov, G.Yu. Bogoslovsky, R.N. Faustov,
V.I. Savrin, L.M. Slad and N.P. Yudin for useful discussions.  The work
was supported by  the grant  UR.02.03.002 of the programme "Universities of
Russia". I.V. was supported in part by the programme SCOPES (Scientific
co-operation between Eastern Europe and Switzerland) of the Swiss National
Science Foundation, project No. 7SUPJ062239, and financed by Federal
Department of Foreign Affairs.

\end{document}